\documentclass[%
 reprint,
superscriptaddress,
nofootinbib,
 amsmath,amssymb,
 aps,
prd,
floatfix,
]{revtex4-2}
\usepackage{graphicx}
\usepackage{subcaption}
\usepackage{dcolumn}
\usepackage{bm}
\usepackage{hyperref}
\usepackage[mathlines]{lineno}
\usepackage[dvipsnames]{xcolor}
\usepackage[normalem]{ulem}

\usepackage{ragged2e}
\usepackage{subcaption}   
\DeclareCaptionJustification{justified}{\justifying}
\newcommand{\change}[1]{#1}

\begin{document}

\title{A Deep Learning Powered Numerical Relativity Surrogate for Binary Black Hole Waveforms}
\author{Osvaldo~Gramaxo Freitas}
\email{osgrade@alumni.uv.es}
\affiliation{Centro de F\'{\i}sica das Universidades do Minho e do Porto (CF-UM-UP), Universidade do Minho, 4710--057 Braga, Portugal}
\affiliation{Departamento de
  Astronom\'{\i}a y Astrof\'{\i}sica, Universitat de Val\`encia,
  Dr. Moliner 50, 46100, Burjassot (Val\`encia), Spain}
  
\author{Anastasios Theodoropoulos}
\affiliation{Departamento de
  Astronom\'{\i}a y Astrof\'{\i}sica, Universitat de Val\`encia,
  Dr. Moliner 50, 46100, Burjassot (Val\`encia), Spain}

\author{Nino Villanueva}
\affiliation{IDAL, Electronic Engineering Department, ETSE-UV, University of Valencia, Avgda. Universitat s/n, 46100 Burjassot, Valencia, Spain}
\affiliation{Departamento de
  Astronom\'{\i}a y Astrof\'{\i}sica, Universitat de Val\`encia,
  Dr. Moliner 50, 46100, Burjassot (Val\`encia), Spain}

\author{Tiago Fernandes}
\affiliation{Centro de F\'{\i}sica das Universidades do Minho e do Porto (CF-UM-UP), Universidade do Minho, 4710--057 Braga, Portugal}
\affiliation{Departamento de
  Astronom\'{\i}a y Astrof\'{\i}sica, Universitat de Val\`encia,
  Dr. Moliner 50, 46100, Burjassot (Val\`encia), Spain}

\author{Solange Nunes}
\affiliation{Centro de F\'{\i}sica das Universidades do Minho e do Porto (CF-UM-UP), Universidade do Minho, 4710--057 Braga, Portugal}

\author{Jos\'e~A.~Font}
\affiliation{Departamento de
  Astronom\'{\i}a y Astrof\'{\i}sica, Universitat de Val\`encia,
  Dr. Moliner 50, 46100, Burjassot (Val\`encia), Spain}
\affiliation{Observatori Astron\`omic, Universitat de Val\`encia,  Catedr\'atico 
  Jos\'e Beltr\'an 2, 46980, Paterna (Val\`encia), Spain}

\author{Antonio~Onofre}
\affiliation{Centro de F\'{\i}sica das Universidades do Minho e do Porto (CF-UM-UP), Universidade do Minho, 4710--057 Braga, Portugal}

\author{Alejandro~\surname{Torres-Forn\'e}}
\affiliation{Departamento de
  Astronom\'{\i}a y Astrof\'{\i}sica, Universitat de Val\`encia,
  Dr. Moliner 50, 46100, Burjassot (Val\`encia), Spain}
\affiliation{Observatori Astron\`omic, Universitat de Val\`encia,  Catedr\'atico 
  Jos\'e Beltr\'an 2, 46980, Paterna (Val\`encia), Spain}

\author{José D. Martin-Guerrero}
\affiliation{IDAL, Electronic Engineering Department, ETSE-UV, University of Valencia, Avgda. Universitat s/n, 46100 Burjassot, Valencia, Spain}
\affiliation{Valencian Graduate School and Research Network of Artificial Intelligence (ValgrAI), Spain}

\newcommand{\Msun}{\mathrm{M}_\odot}
\newcommand\e[1]{{\times}10^{#1}}
\newcommand{\targsur}{\texttt{NRHybSur3dq8}}
\newcommand{\dansur}{\texttt{DANSur3dq8}}

\date{\today}

\begin{abstract}
Gravitational-wave (GW) approximants are essential for gravitational-wave astronomy, allowing the coverage of the binary black hole parameter space for inference or match filtering without costly numerical relativity (NR) simulations, but generally trading some accuracy for computational efficiency. To reduce this trade-off, NR surrogate models can be constructed using interpolation within NR waveform space. We present a 2-stage training approach for neural network-based NR surrogate models. \change{We initially train four models on waveforms generated from four different GW approximants, and then fine-tune these models on NR data. We show that despite the median mismatches of the pre-trained models with NR ranging over two orders of magnitude, the fine-tuned models all reach median mismatches of order $10^{-5}$, on par with top-performing NR surrogates. The dual-stage artificial neural surrogate (\dansur) models also offer rapid waveform generation, with millions of waveforms being generated in under 20ms on a GPU.} Implemented in the \textsc{bilby} framework, we show \dansur~ can be used for parameter estimation tasks.


\end{abstract}


\maketitle

\section{\label{sec:intro} Introduction}
Gravitational-wave (GW) astronomy is a computationally intensive field. Within the LIGO-Virgo-KAGRA (LVK) collaboration~\cite{aligo_2015,avirgo_2015,kagra_2019}, a number of detection pipelines  (GstLAL~\cite{sachdev2019gstlal}, MBTA~\cite{Aubin_2021}, PyCBC~\cite{Dal_Canton_2021}, SPIIR~\cite{chu2021spiir}, and cWB~\cite{Klimenko_2016}) are constantly monitoring detector data in order to inform collaboration members of potential GW signatures, as well as facilitate public alerts for the broader astrophysical community. If a GW trigger is detected by the pipelines of the LVK collaboration, a further analysis is often desired for parameter estimation. In the case of binary black hole (BBH) events, this is done by maximizing some likelihood function based on a similarity metric between the detector data and simulated GW waveforms, typically a noise-weighted inner product, which should reach a maximum if the original wave is exactly the same as some produced simulation~\cite{Maggiore_book}. 
This means one must have at hand some process which, given some physical parameters as input, will produce a physically accurate waveform, while being fast enough to allow an exploration of the GW parameter space. Waveforms from full numerical relativity simulations, which can take weeks to complete, are not appropriate for this task. Instead, one typically uses~\textit{approximants}, computational routines which produce fast approximations of what a gravitational waveform should look like given some physical parameters. As they are ultimately tested against (and often calibrated according to) numerical relativity waveform catalogues (such as the ones produced by the SXS collaboration~\cite{SXS}), approximants tend to be highly accurate and indeed have allowed for the analysis of all GW signals detected to date~\cite{GWTC-1,GWTC-2,GWTC-2.1,GWTC-3}. 

Despite this remarkable success, there is still room for improvement in the field. Parameter estimation methods relying on repeated waveform evaluations, such as \textsc{bilby}'s Bayesian inference with dynamic nested sampling~\cite{bilby_paper}, benefit greatly from fast waveform generation. This process can be parallelized in order to speed up inference times (note, however, that this speedup is not linear with the number of cores~\cite{dynesty_scaling}). This is usually done by sending Bayesian inference jobs to high-performance computing (HPC) clusters, making use of large numbers of CPU cores to obtain the desired parallelization gains. However, even for these implementations the computational bottleneck can be found in the calculation of the likelihood, in particular the generation of the waveform. Methods like \textsc{RIFT} can also benefit from fast waveform generation, though to a lesser extent, as they require a single step of waveform generation over a pre-defined grid \cite{RIFT}.

Next-generation GW detectors, namely Einstein Telescope and Cosmic Explorer, are expected to achieve detection rates of 100,000 BBH events per year~\cite{ET_rates,CE_rates}. In order to process these events efficiently with Bayesian methods, the waveform generation process must be made faster, while still allowing for an accurate parameter estimation. In order to address this issue, applications of deep learning methods to steps of this process have been proposed. Approaches based on likelihood-free inference avoid live waveform generation altogether by replacing Bayesian parameter estimation with neural posterior estimation methods (though efficient waveform generators are still useful for the training process)~\cite{dingo_paper,PERCIVAL}, while others strive for a speed-up through more efficient sampling using normalizing flows~\cite{nessai,nessai_sw}. 

In this paper, we instead tackle the problem through the lens of speeding up waveform generation. Existing literature has shown machine learning to be an effective method for waveform generation, in particular to provide reduced order models of existing approximants. Previous works have successfully demonstrated the generation of accurate waveforms through the mapping of physical parameters to the coefficients of a reduced order basis obtained through greedy algorithms~\cite{ROMAN, ANN_Sur_FD, LucyThomasANN,Nousi_spiral}, an approach which has been reproduced for PCA bases~\cite{schmidt_ml_gw_bbh,grimbergen2024generatinghigherordermodes}, as well as SVD bases ~\cite{Luna2024,fontbute2024,SVD_meshfree}. 

We expand on previous works by introducing a strategy for the training of fast and accurate neural surrogate models for numerical relativity data. By training a small neural network to predict the coefficients of a reduced PCA basis of approximant waveforms, and then fine-tuning on numerical relativity waveforms, allowing the network to dynamically adjust the basis, it is possible to obtain a model which is capable of generating millions of accurate waveforms on a consumer-level GPU in a matter of milliseconds. As a demonstration of this training strategy, we present ~\dansur~ (\textbf{D}ual-stage \textbf{A}rtificial \textbf{N}eural \textbf{Sur}rogate), a fast and accurate surrogate to the dominant (2,2) mode of numerical-relativity-produced GWs leveraging the GPU-based nature of modern machine learning. \change{As a show of generalizability, we produce 4 versions of~\dansur, pre-training on datasets obtained from 4 different GW approximants.}

The paper is organized as follows: in Section~\ref{sec:dataset}, we introduce the datasets used in this work, including the different approximants and numerical relativity waveforms. Section~\ref{sec:methods} describes our methodology, detailing the dimensionality reduction approach, neural network architecture used for training and the training workflow. The results of our model, including accuracy, speed, and parameter estimation capabilities, are presented in Section~\ref{sec:res}\@. Finally, we summarize our findings and discuss potential future directions in Section~\ref{sec:conclusions}.

The  code developed to create and train the models used in this work can be found at \url{https://github.com/osvaldogramaxo/DANSur_22/}.

\section{\label{sec:dataset} Datasets}
In the development of deep learning methods, data availability often poses a significant constraint. This proves particularly problematic in the field of numerical relativity, where the high costs associated with simulations result in a scarcity of data. For instance, by applying constraints of an effective precessing spin $\chi_p < 0.001.$ and a mass ratio $q < =8$ to the SXS BBH data~\cite{SXS}, we obtain only $376$ samples of numerical relativity simulations. Given that deep learning frameworks typicaly need very large datasets in order to properly interpolate between the points of the input and output parameter spaces,  current numerical catalogues are inadequate to fully explore them.
To compensate for this, we propose to generate a substantial GW dataset using a existing approximants to train a model which can capture the fundamental characteristics of BBH GW signals, and subsequently fine-tune this model using numerical relativity data. \change{In our study we will use four approximants: \texttt{NRHybSur3dq8}, a best-in-class surrogate model built using an empirical interpolant~\cite{NRSur7dq8ref}, \texttt{SEOBNRv5HM}, an approximant making use of the effective-one-body (EOB) formalism in order to generate GW waveforms~\cite{SEOBNRv5HM}, \texttt{SEOBNRv4HM\_PA}, an EOB approximant accelerated by using a post-adiabatic approach to evolve the system dynamics~\cite{SEOBNRv4HM_PA}, and \texttt{IMRPhenomTHM}, an approximant making use of phenomenological fits to generate waveforms in the time domain~\cite{IMRPhenomTHM}. 
Data for the latter two is generated through their \textsc{lalsimulation} implementation \cite{lalsuite}, data for \texttt{NRHybSur3dq8} is generated using the ~\textsc{gwsurrogate} package for python~\cite{gwsurrogate_ref}, and data for \texttt{SEOBNRv5HM} is generated with the \textsc{pyseobnr} package \cite{pyseobnr}.
}
\subsection{Approximant datasets}
\begin{figure}
\centering

\includegraphics[width=1\columnwidth]{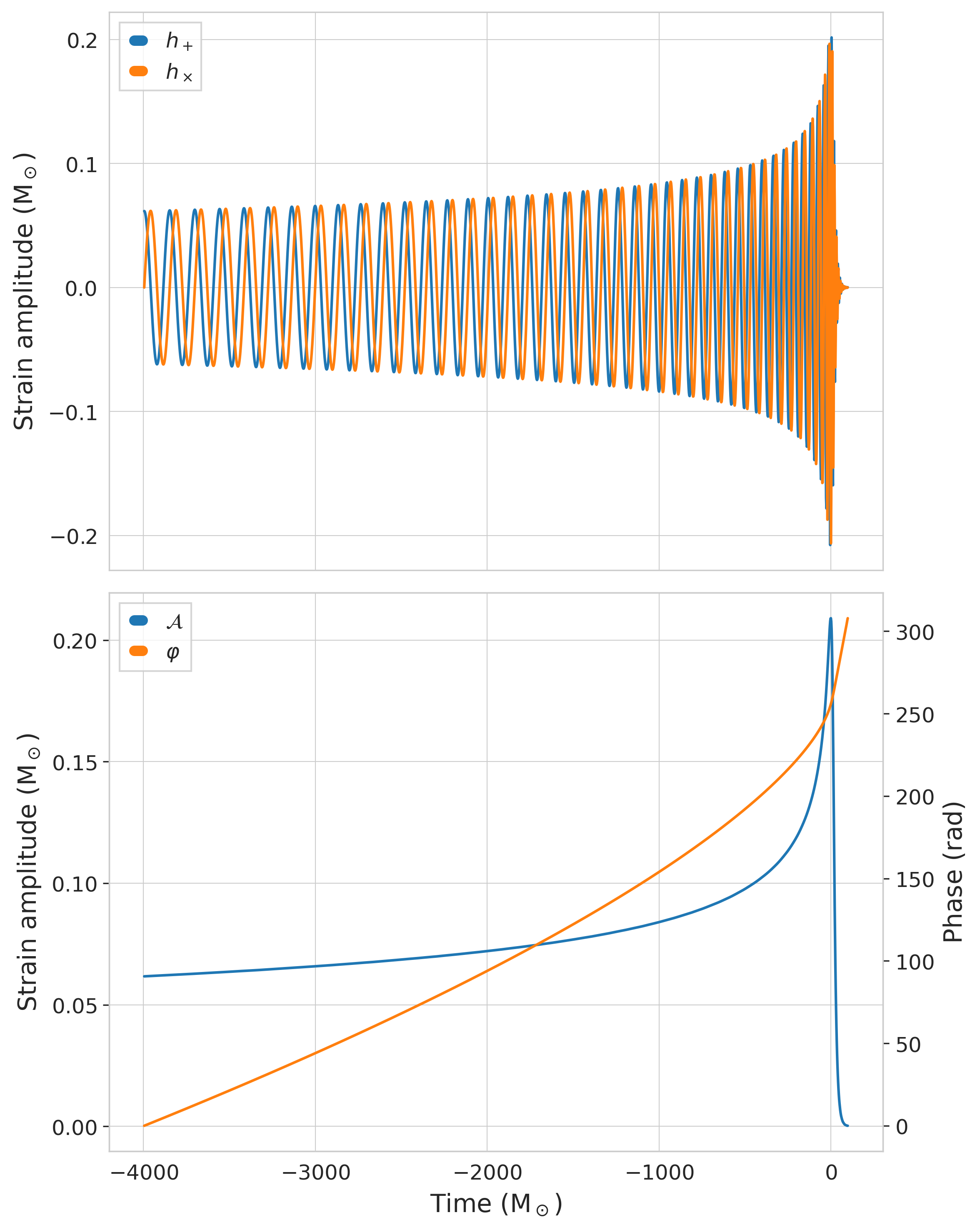}

\caption{\label{fig:example} An example of a BBH waveform generated using the approximant, in the polarization (top) and amplitude-phase (bottom) representations.}
\end{figure}

\change{For each approximant}, we generate 102,400 waveforms. We restrict ourselves to aligned-spin BBH mergers, in order to keep complexity manageable, with mass ratios sampled uniformly from the $[1, 8]$ interval, while the dimensionless component spins $\chi_1,\chi_2$ are sampled uniformly from the $[-0.8,0.8]$ interval. 
We note that while uniform sampling does not necessarily reflect population-informed priors, it avoids biasing towards any particular parameter combination. The waveforms are generated in simulation units, with amplitude and time scaled by the sum of the Christodoulou masses of both black holes, with a corresponding sample rate of $2~\Msun^{-1}$. The merger time is estimated as the time which maximizes the amplitude of the quadrupolar $(l,|m|)= (2,2)$ mode, and the waveforms are trimmed to a duration of $4096~\Msun$, lining up the merger $100~\Msun$ before the end of the window. We further simplify our problem by setting the initial phase of the (2,2) mode in all the waveforms to zero. The (2,2) mode of the waveform is then saved to an HDF5 file, along with the physical parameters for the generation.

\subsection{Numerical Relativity waveforms}
For setting up the numerical relativity dataset, we make use of the publicly available BBH simulation data from the SXS collaboration~\cite{SXS}, consisting of 2018 simulations. The (2,2) waveforms are extracted using the SXS python package~\cite{sxs_package_2024}. To match the data we produced with the approximant, we must do a number of checks. First, we take into account only the waveforms which have a duration of at least $4096~\Msun$. Then, we select the non-precessing waveforms by filtering effective precessing spins $\chi_p<0.001$~\cite{eff_prec_spin}, with mass ratios $q\leq8$, so that we are in line with the approximant data described previously. \change{We also make sure memory effects are not present}. This process leaves us with only 376 waveforms. These waveforms are interpolated using cubic splines to the same sample rate as the approximant dataset, and trimmed to $4096~\Msun$.

\section{\label{sec:methods} Methods}

\subsection{\label{sec:dimensionality_reduction}Dimensionality reduction}
When dealing with numerical data, gravitational waveform modes are essentially arrays of $M$-length complex numbers. Formally, these are points in $\mathbb{C}^M$, which can be reframed as points in $\mathbb{R}^{N}$ with $N=2M$. However, the high correlation between adjacent points in the waveform, as well as restrictions to the shape of the waveforms (that is, wavelike signals with a chirp and ringdown-like features) means that gravitational waveforms occupy a very restricted subspace of  $\mathbb{R}^{N}$. On the other hand, neural networks with $N$ outputs will output to some generic subspace of $\mathbb{R}^{N}$, the size of which will depend on the number of layers and neurons per layer (that is, its expressivity). Therefore, we need the network to learn not only the shape of the subspace, as well as how this space is connected. Furthermore, the average Euclidean distance between two points in a cube of side $l$ in  $\mathbb{R}^{N}$ scales as $O(\sqrt{N})$~\cite{avg_dist_ndim_ref} . 

 This means that, in general, high dimensional data will be more sparse than lower dimensional data. As such, a neural network will have to interpolate over longer distances in such a case, which may lead to problems with convergence and generalisation~\cite{curse_of_dim}. Therefore, the projection of $N$-dimensional data to some lower-dimensional space should make the data more dense and thus facilitate the ease of convergence of neural networks training on that data. This is the concept of dimensionality reduction, and it is widely used across applications of machine learning~\cite{DimRed_grapes,DimRed_phys,DimRed_wav}. 
There are many methods for dimensionality reduction, both linear and non-linear. Our use case involves training a model in a lower-dimensional representation space of gravitational waves, and then performing an inverse transform from that space back to the typical waveform representation. This requires that any method we choose must have an accurate inverse transform. While linear methods typically have an easily calculable inverse, non-linear methods very often do not have a tractable inverse, and they must be fitted to the data in a similar way as the forward transform, such as in autoencoders or UMAP~\cite{ae_ref,umap}. In this work, we will focus on principal component analysis (PCA), one of the most successful linear dimensionality reduction methods~\cite{PCA_ref,pca_rev}. It is worth noting that SVD works similarly to PCA and as such it would be expected to yield similar results. Similarly, other dimensionality reduction methods such as the aforementioned greedy basis methods can be used in a similar way. \change{While a thorough comparison of dimensionality reduction methods for the creation of surrogate models is lacking in literature, the issue has been discussed in theory at length in \cite{ROMs_Surs_review}, while a comparison of SVD and greedy basis methods can be found in appendix B of \cite{svd_vs_greedy}. }

\begin{figure}
  \centering
  \makebox[0pt]{\includegraphics[width=0.5\textwidth]{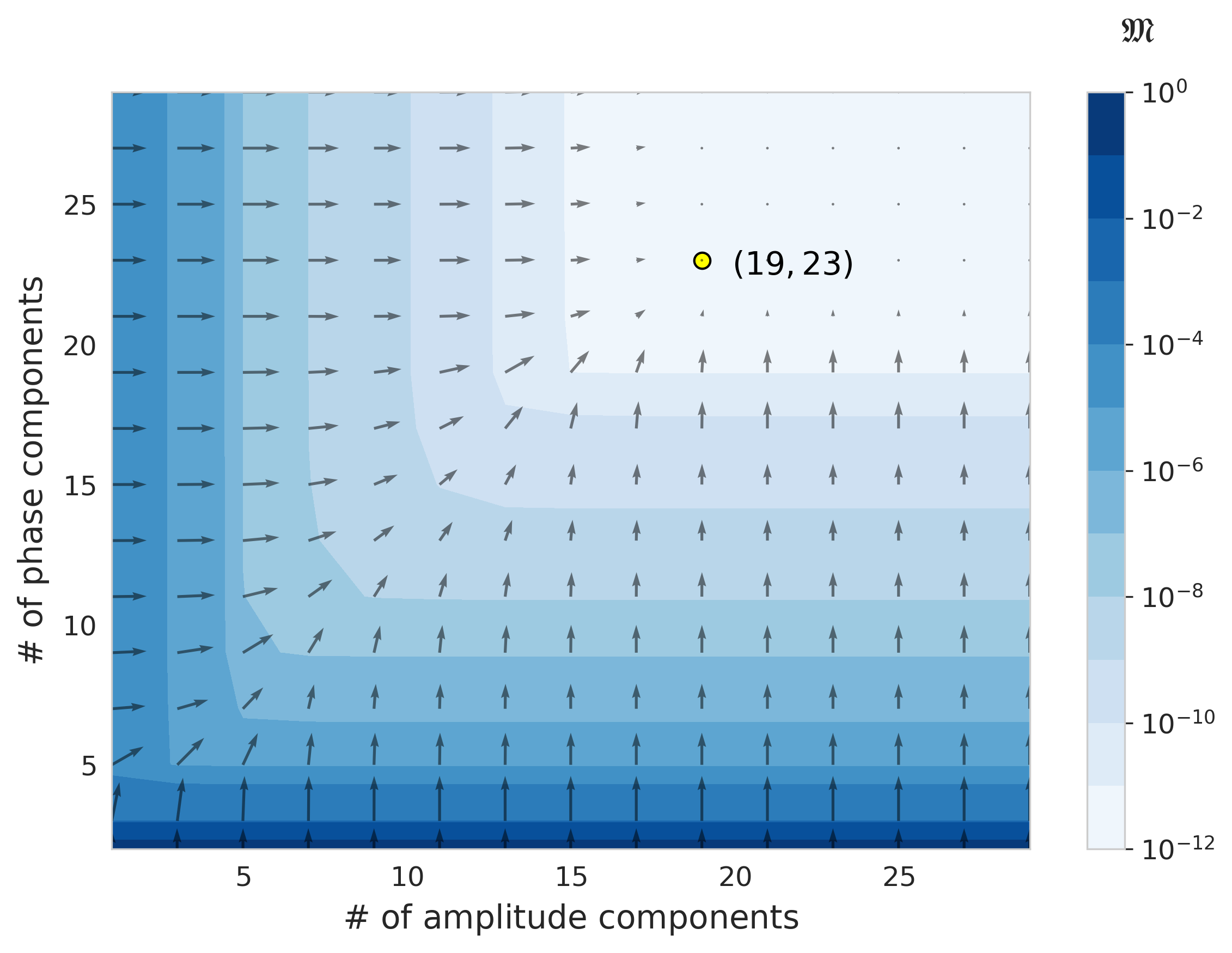}}
  \caption{\label{fig:phase_pca} Mean mismatch values as a function of the number of principal components kept for the amplitude and phase. The arrows show the gradient of the mismatch values, and the yellow circle shows the chosen number of components to build the PCA basis.}
\end{figure}
PCA works by calculating a linear transform $\mathcal{L}: X~\rightarrow (X-\mu)V$, where $X$ is some input $N\times M$ matrix (interpreted in our case as being a collection of $N$ waveforms of length $M$),  $\mu$ is the average value at each timestep, and $V$ is the $m\times n$ matrix of eigenvectors of the covariance matrix $\Sigma=\frac{1}{n}(X-\mu)^T(X-\mu)$. The crucial point here is that the centering of $X$ means that the eigenvalues of $\Sigma$ represent the explained variance of the corresponding eigenvector. As such, if we consider only the eigenvalue-sorted subset of the top-valued $k$ eigenvectors, $V_k$  (an $m\times k$ matrix), we can project the data onto the reduced basis of dimension $k$ through the PCA transform:
\begin{eqnarray}
\label{eq:pca_t}
X^{PCA}_K=(X-\mu)V_k.
\end{eqnarray}
We can then approximately reconstruct the original dataset through the inverse PCA transform
\begin{eqnarray}
\label{eq:invpca_t}
X'=X^{PCA}_K V^T_k+\mu\,.
\end{eqnarray}
There will be some error $E(X,X')$ associated to this reconstruction, which will depend on the amount of components truncated (that is, $E(X,X')\rightarrow 0$ as $k~\rightarrow n$).

In our case, we will separate the complex waveform arrays $h = h_+ - ih_\times = Ae^{i\phi}$ into 2-channel arrays, separating the amplitude $\mathcal{A}$ and (unwrapped) phase $\varphi$ quantities, as represented in Figure~\ref{fig:example}. This allows for a simpler, numerically non-oscillating representation, and also allows us to perform PCA separately on each quantity. Therefore, we can tune the amount of precision required in the reconstruction of each quantity. \change{In Figure~\ref{fig:phase_pca} we show the average mismatch values (see definition below) and their gradients as a function of the number of principal components kept for the amplitude and phase, calculated for the ~\targsur~dataset. As can be seen, the influence of the number of phase components slightly outweighs the amplitude, and after 19 amplitude components and 23 phase components, the gradients essentially vanish. Going forward, our chosen representation will thus keep 19 components for the amplitude and 23 components for the phase, thus reducing the 2048-dimensional complex representation of waveforms (which is equivalent to a 4096-dimensional representation) to a 42-dimensional one.}

\subsection{Network architecture and training}
Conceptually, we propose to build a neural network that maps the 3 physical input parameters (mass ratio and dimensionless individual spins) to the 42-dimensional orthogonal space corresponding to the chosen PCA bases. We can then apply the inverse PCA transform and obtain the reconstructed waveform $h$. With this formulation, we can now define our optimization objective, or loss function. On the one hand, we want to ensure the accuracy of the predicted PCA components. This can be quantified using a mean absolute error (MAE) criterion,
\begin{equation} 
L_1(c,~\hat{c}) =~\frac{1}{N}~\sum_{i=1}^{N}~\left| c_i -~\hat{c}_i~\right|, 
\label{eq} 
\end{equation}
where $c$ is the generated array of the N PCA components (N=42 in our case) for each generated waveforms, and $\hat{c}$ represents the corresponding ground truth values of these components. We choose to use MAE rather than its square as it is more robust to outliers.
On the other hand, from a practical perspective, we eventually want to maximize the overlap $\mathcal{O}$ between the waveforms themselves, as that is the usual evaluation metric for approximants. It should be useful, then, to make sure our loss function includes information about the overlap. Note that since we want to~\textit{maximize} the overlap, we should then~\textit{minimize} the mismatch $\mathfrak{M}$, given by 

\begin{equation}
  L_2 (h,~\hat{h}) = 1 -~\mathcal{O}(h,\hat{h}) =1-~\frac{\langle h |~\hat{h}~\rangle}{\sqrt{\langle h | h~\rangle~\langle~\hat{h} |~\hat{h}~\rangle}},
 ~\label{eq:overlap_gravitational_waveforms}
\end{equation}
where~\[
\langle h |~\hat{h}~\rangle = 4\,\text{Re}\!\!\int_{-\infty}^{\infty}\!\!\!\! \tilde{h}(f)\tilde{\hat{h}}^*(f) df \approx 4\,\text{Re}\sum_{k}\tilde{h}_k\tilde{\hat{h}}_k^*\Delta f,
\] is the inner product between waveforms (note that if this operation is being applied to compare a waveform with detector data instead, the integral limits should change to the relevant frequency range). \change{While the mismatch is often optimized over time and phase, the alignment of the waveforms and the setting of the initial phase in our datasets make this unnecessary.} Numerically, it will be more convenient to work with the base-10 logarithm of this quantity, as this will keep the magnitude of the loss (and gradients) more consistent as the mismatch reaches lower values. As such, our loss function $\mathcal{L}$ shall be
\begin{equation}
   ~\mathcal{L} = L_1 +~\log(L_2)\,.
   ~\label{eq:placeholder_label}
\end{equation}
\begin{figure}
\includegraphics[width=.4\columnwidth]{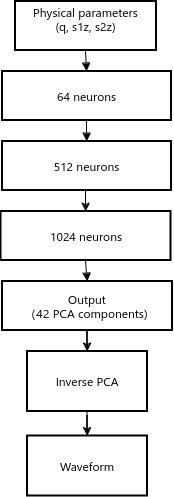}
\caption{\label{fig:model} Visualisation of model architecture.}
\end{figure}
\begin{figure}
  \centering
  \makebox[0pt]{\includegraphics[width=0.5\textwidth]{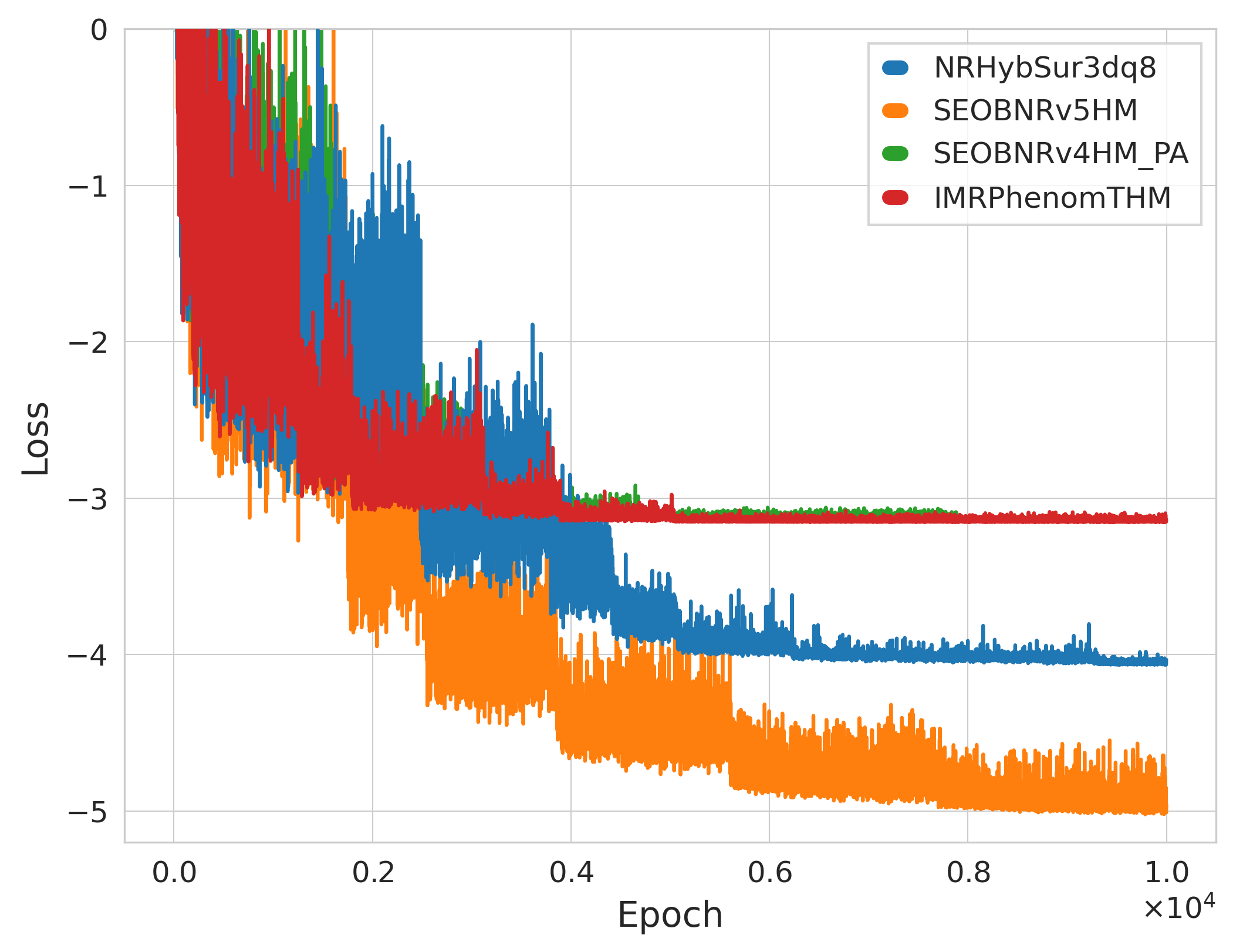}}
  \caption{\label{fig:approx_loss} Validation loss on the approximant datasets during pre-training.}

\end{figure}

The network used is a simple multi-layered perceptron with 3 hidden layers, with 64, 512 and 1024 neurons respectively, using GELU activation functions. The model architecture is illustrated in Figure~\ref{fig:model}. In this work, all computations make use of the NVIDIA H100 GPU nodes provided by the Artemisa cluster at the University of València. The number and width of the hidden layers are the result of an optimization process through the~\textsc{optuna} package~\cite{optuna_2019} (note that in the case of the layer width, we optimized over powers of $2$, in order to reduce redundancy). The model is then pre-trained for \change{up to 10,000 epochs for each} approximant dataset, with an 80/20 train/validation split. We use the Sophia optimizer~\cite{sophiaG}, with an initial learning rate obtained through the learning rate finder process outlined in~\cite{lrfinder}, and a learning rate scheduler reducing the learning rate by a factor of 2 after 500 epochs without improved performance on the validation set. During the training loop we monitor the mean value of the validation mismatches, and save the model every time there is an improvement (that is, when the mean validation mismatch decreases in comparison to the previous best value), guaranteeing we have the best performing model on the training set after all epochs are done. This model will then be fine-tuned on the numerical relativity data. Since the numerical relativity dataset is quite small, we proceed as follows: 
first, we set apart 1/8th of the waveforms for testing purposes. 
With the remaining waveforms, we use k-fold cross validation by splitting the data in 5 equal-sized subsets or ``folds''. For each iteration, we consider the non-chosen folds as a training set, and the chosen fold as the validation. This allows us to have 5 iterations of training, using the model trained in the approximant as a starting point, and using a lower learning rate than in the pre-training step ($1\times10^{-4}$) in order to avoid an excessive deviation from the region of the space of network weights arrived at in pre-training. When this process is completed, the weights of the 5 trained models are averaged to obtain a final model, which is then evaluated on the test set. This process should be seen as a sort of weight averaging, which has been shown to lead to improved generalization performance~\cite{weight_avg}.
\begin{figure}
    \centering
    \makebox[0pt]{\includegraphics[width=0.5\textwidth]{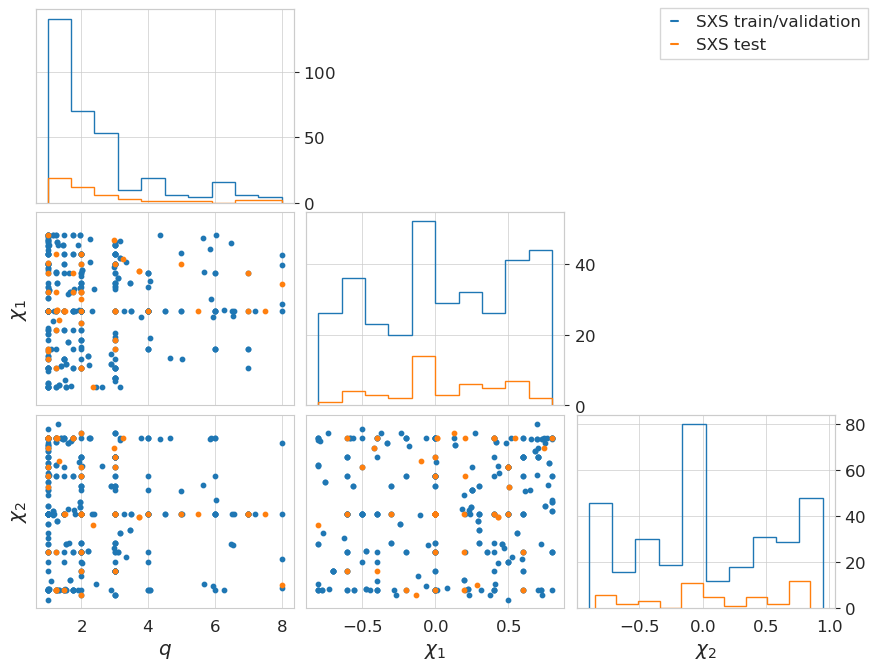}}
    \caption{Distribution of parameters in the NR dataset's train/validation and testing sets.}
    \label{fig:sxsdist}
\end{figure}

\begin{figure}
\makebox[0pt]{\includegraphics[width=0.5\textwidth]{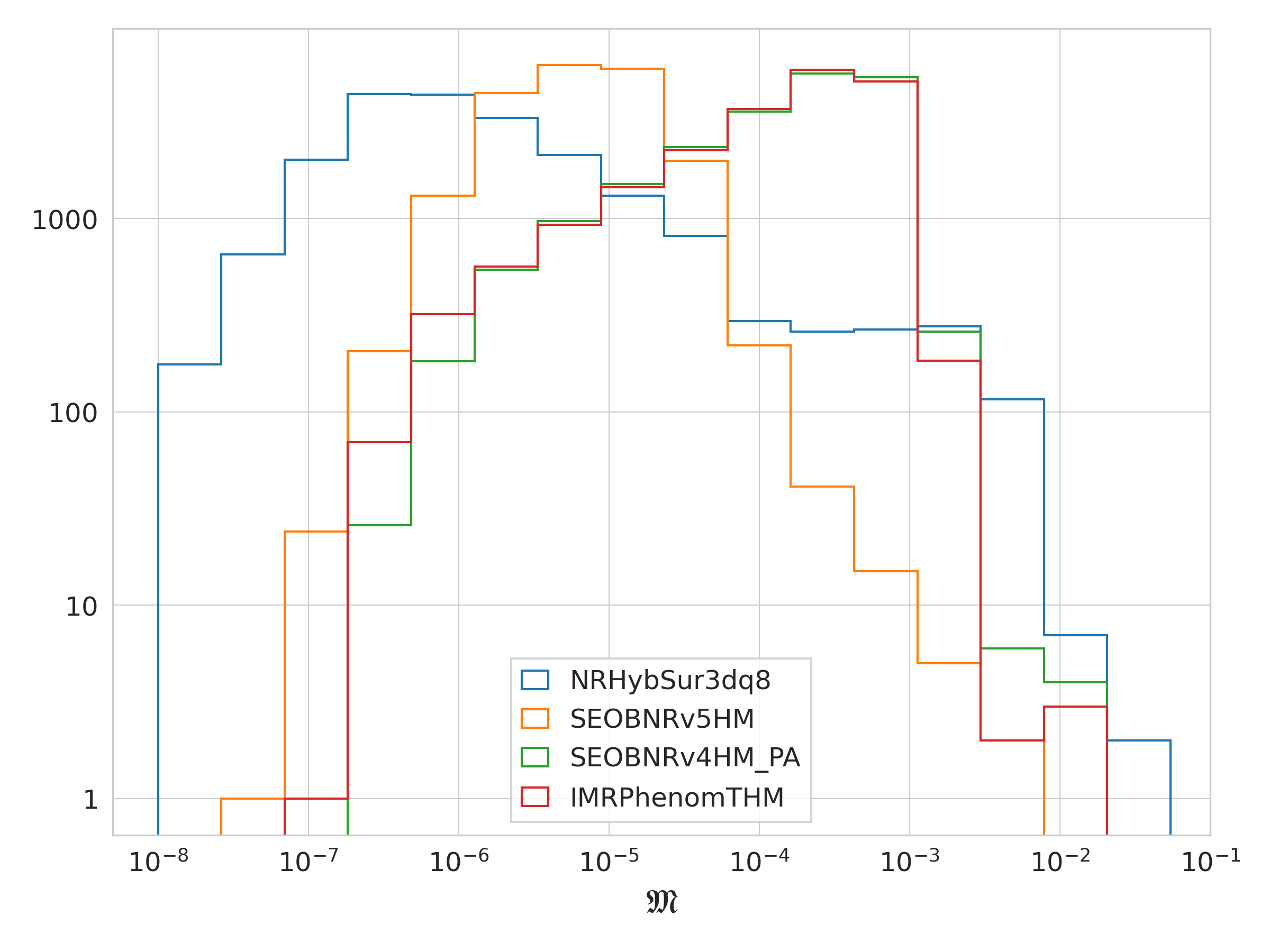}}
\caption{\label{fig:mms} Distribution of validation mismatches for the models trained on the approximant datasets.}
\end{figure}
\begin{figure}
\makebox[0pt]{\includegraphics[width=0.5\textwidth]{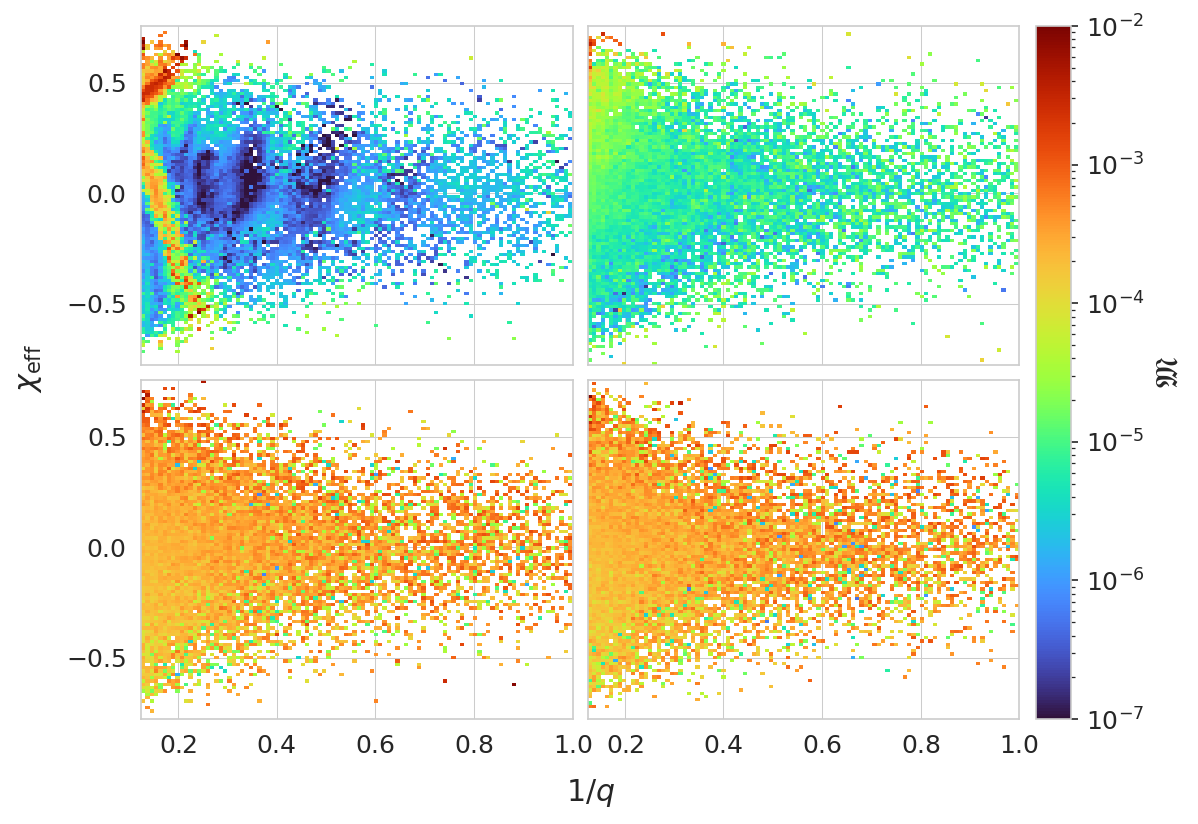}}

\caption{\label{fig:mms_scatter} Heatmaps showing the mean validation mismatch of the pre-trained network per bin, as a function of the (inverted) mass ratio and the effective inspiral spin. \textbf{Top left:}  \texttt{NRHybSur3dq8}. \textbf{Top right:}  \texttt{SEOBNRv5HM}.  \textbf{Bottom left:}  \texttt{SEOBNRv4HM\_PA}.  \textbf{Bottom right:}  \texttt{IMRPhenomTHM}.}

\end{figure}

\section{\label{sec:res} Results}

\subsection{Accuracy}
\subsubsection{Results on the approximant waveforms}

\change{In Figure~\ref{fig:mms} we depict the mismatch distributions for each of the four individual waveform approximants. The \texttt{NRHybSur3dq8} model yields a broad distribution, featuring a significant population with mismatches below $10^{-6}$ but also a flat feature near $10^{-4}$, with a tail extending to $2.5\e{-2}$. The distribution for \texttt{SEOBNRv5HM} peaks around $10^{-5}$, with its worst mismatches reaching $3.4\e{-3}$. Lastly, the \texttt{SEOBNRv4HM\_PA} and \texttt{IMRPhenomTHM} models display very similar and more concentrated distributions, both showing a ramp up to the peak around $10^{-3}$, with a sharp drop off to maximum mismatches of order $10^{-2}$.}
\change{Figure~\ref{fig:mms_scatter} complements this information by showing the mean mismatch value per bin over the plane defined by the inverted mass ratio and the effective inspiral spin $\chi_{\rm eff}=\frac{\chi_1 +q\chi_2}{1+q}$ for each of the approximants. The similarity between \texttt{SEOBNRv4HM\_PA} and \texttt{IMRPhenomTHM} is once again made evident, with both showing mostly uniform mean values between $10^{-4}$ and $10^{-3}$. Given that both these approximants are used through their \textsc{LALSimulation} implementation, it is possible that there is some common processing in the generation of these waveforms. The network trained on the \texttt{SEOBNRv5HM} approximant data also shows fairly uniform behaviour across the parameter space, with some higher mismatches appearing at higher $\chi_{\rm eff}$ values, but otherwise remaining between $10^{-6}$ and $10^{-4}$. The behaviour on the \texttt{NRHybSur3dq8} data stands out as most interesting, with the appearance of some strong heterogeneity over the parameter space, with particular expression near the origin, where a diagonal line from $(\frac{1}{q}=0.25,\chi_{\rm eff} = -0.7)$ to $(\frac{1}{q}=0.125,\chi_{\rm eff} = 0.3$) shows mean mismatches up to three orders of magnitude larger than its surrounding regions. It is unclear why this happens and visual inspection of the generated waveforms does not show any collapse behavior. Nonetheless, even in these cases the mismatches are decent and we do not expect it to affect the fine-tuning process significantly.}

\subsubsection{Results on the Numerical Relativity waveforms}

\begin{figure*}
\centering
\makebox[0pt]{\includegraphics[width=1\textwidth]{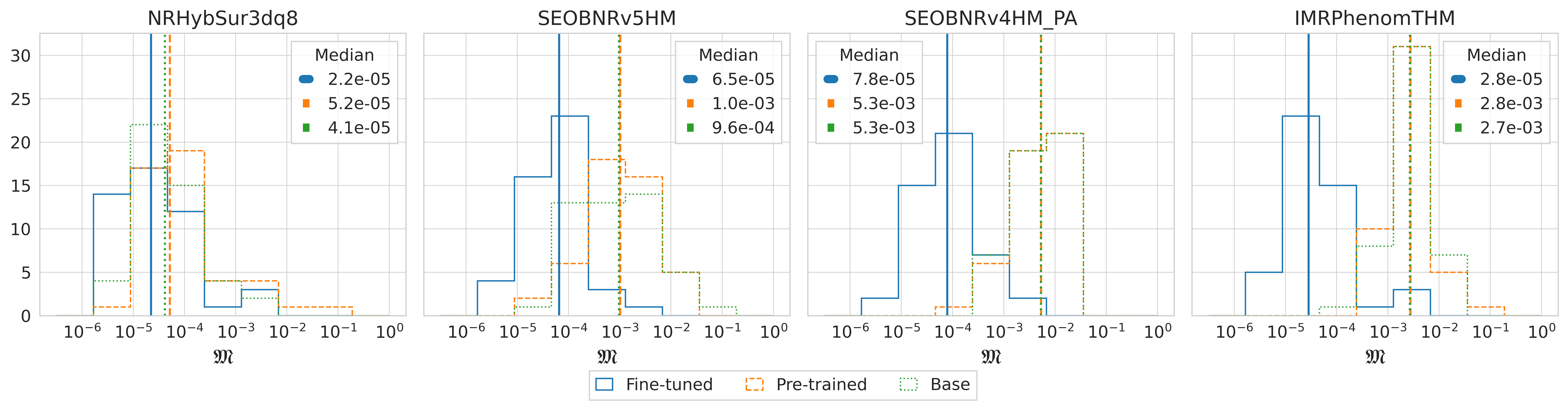} }
\caption{\label{fig:mms_NR} Distribution of mismatches on the set-aside test NR dataset, shown for the fine-tuned and weight-averaged models, as well as for the pre-trained-only models, along with the mismatches of the base approximants themselves with NR.}

\end{figure*}

\change{Table~\ref{tab:table_label} shows the mean and maximum mismatch values for the various trained models in the set-aside NR test set. We can see that the weight averaging of the 3 models trained on the folds seems to achieve the goal of enforcing generalization capacity. While not always the top performing model, the weight-averaged models always avoid having large maximal mismatches, while retaining mean performance. This lines up with the theoretical advantages of weight-averaging regarding model generalization. In Figure~\ref{fig:mms_NR} we show the mismatches for the fine-tuned models' reconstructions of NR test set waveforms, when pre-trained on different datasets. We also show the mismatches of the corresponding model after the pre-training step, as well as the mismatches obtained using the base approximant used to train the model. We should start by pointing out the difference in accuracy to NR across the different approximants, with the \texttt{NRHybSur3dq8} approximant achieving the best median mismatches, around $4\e{-5}$, and the \texttt{SEOBNRv4HM\_PA} approximant being the relatively worst performer with median mismatches of $5.3\e{-3}$, two orders of magnitude larger than the best performer. In all cases, the median mismatches of the pre-trained models correspond remarkably well with the mismatches of the respective base approximants, showing that the proposed architecture can accurately reproduce the outputs of GW approximants. The fine-tuning process yields very different results depending on the base approximant used for pretraining. In the \texttt{NRHybSur3dq8} case, the major gain with the fine-tuning process is the suppression of a tail that previously extended beyond $10^{-1}$, arriving at a final distribution with a median of $2.2\e{-5}$, smaller than the median from the approximant by almost a factor of 2. For the other approximants, however, the fine-tuning process shows gains in the median mismatch between one and two orders of magnitude, with every fine-tuned model achieving median mismatches below $8\e{-5}$, matching the accuracy from the base \texttt{NRHybSur3dq8} approximant, and in the case of the model pre-trained on \texttt{IMRPhenomTHM}, surpassing it slightly. The convergence to this range of values regardless of the accuracy of the base approximant suggests that the strategy used to train \dansur~ can successfully use synthetic data obtained from approximations as a springboard for learning to reproduce simulation data, with accuracy gains that can span multiple orders of magnitude over the initial approximation.}

\renewcommand{\arraystretch}{1.3}
\begin{table}
\centering
\begin{tabular}{llcc}
\hline
\textbf{Model} & \textbf{Config} & \textbf{Mean} & \textbf{Max} \\
\hline
\hline
\texttt{NRHybSur3dq8} 
 & W. Avg.   & $2.3 \times 10^{-5}$ & $2.7 \times 10^{-3}$ \\
 & Fold 1 & $2.8 \times 10^{-5}$ & $\mathbf{2.0 \times 10^{-3}}$ \\
 & Fold 2 & $\mathbf{2.0 \times 10^{-5}}$ & $1.0 \times 10^{-2}$ \\
 & Fold 3 & $3.3 \times 10^{-5}$ & $3.0 \times 10^{-3}$ \\
\hline
\texttt{SEOBNRv5HM} 
 & W. Avg.   & $5.3 \times 10^{-5}$ & $\mathbf{2.8 \times 10^{-3}}$ \\
 & Fold 1 & $\mathbf{4.7 \times 10^{-5}}$ & $1.0 \times 10^{-2}$ \\
 & Fold 2 & $5.8 \times 10^{-5}$ & $5.8 \times 10^{-3}$ \\
 & Fold 3 & $8.3 \times 10^{-5}$ & $5.7 \times 10^{-3}$ \\
\hline
\texttt{SEOBNRv4HM\_PA} 
 & W. Avg.   & $6.2 \times 10^{-5}$ & $\mathbf{6.4 \times 10^{-3}}$ \\
 & Fold 1 & $\mathbf{4.9 \times 10^{-5}}$ & $4.0 \times 10^{-2}$ \\
 & Fold 2 & $7.5 \times 10^{-5}$ & $1.3 \times 10^{-2}$ \\
 & Fold 3 & $6.0 \times 10^{-5}$ & $6.6 \times 10^{-3}$ \\
\hline
\texttt{IMRPhenomTHM} 
 & W. Avg.   & $\mathbf{2.7 \times 10^{-5}}$ & $4.9 \times 10^{-3}$ \\
 & Fold 1 & $3.5 \times 10^{-5}$ & $1.4 \times 10^{-2}$ \\
 & Fold 2 & $3.8 \times 10^{-5}$ & $2.2 \times 10^{-2}$ \\
 & Fold 3 & $3.2 \times 10^{-5}$ & $\mathbf{2.7 \times 10^{-3}}$ \\
\hline
\end{tabular}
\caption{Mean and maximum mismatch values for the various models in the set-aside test set, showing results for individual folds and the weight-averaged model\label{tab:table_label}}

\end{table}
\subsection{Speed}

\begin{figure}
 \centering
\makebox[0pt]{\includegraphics[width=0.5\textwidth]{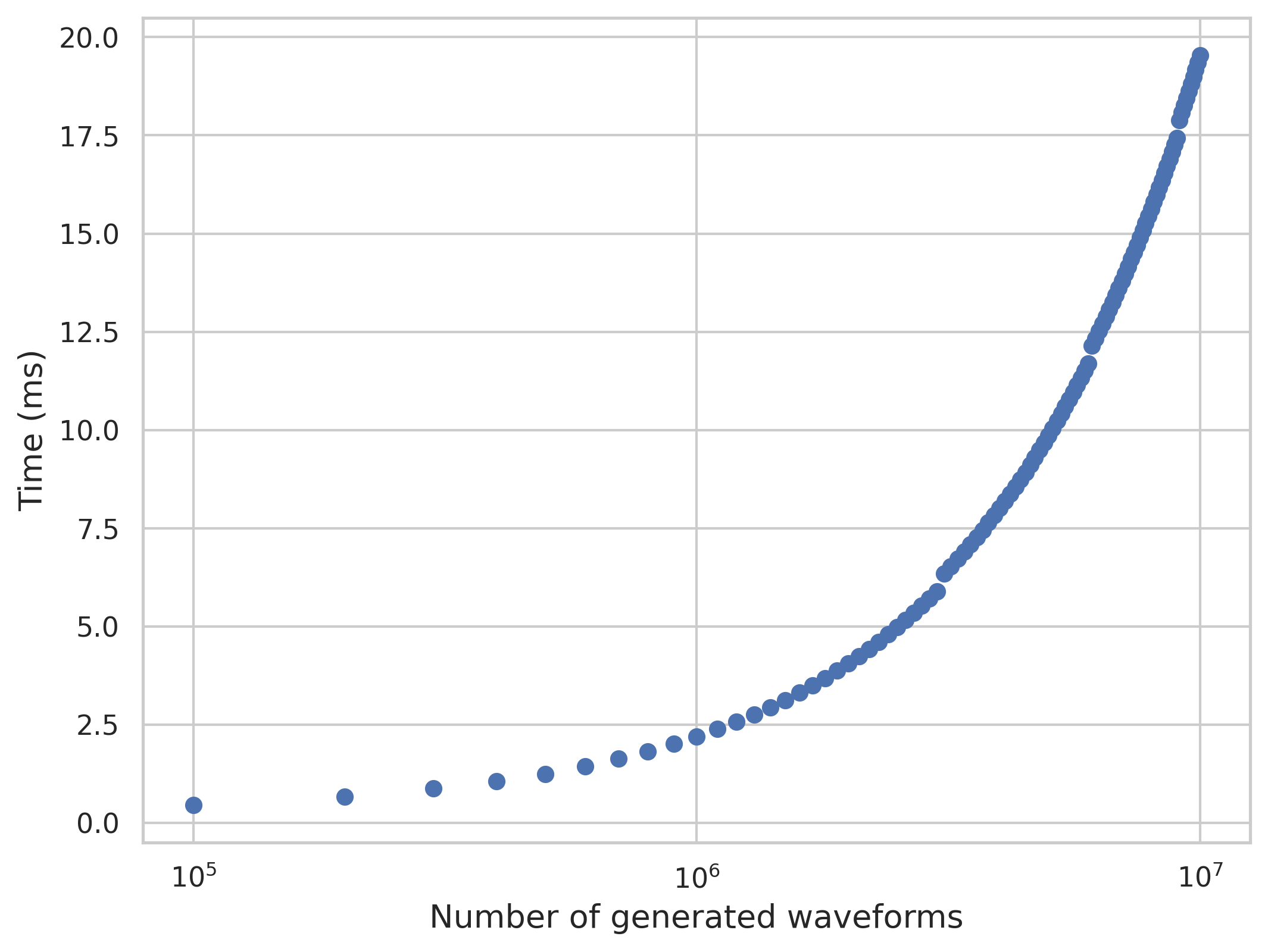}}
 \caption{\label{fig:t_cumul}Cumulative waveform generation time for~\dansur.}
 
\end{figure}
\begin{figure}
 \centering
 \includegraphics[width=\linewidth]{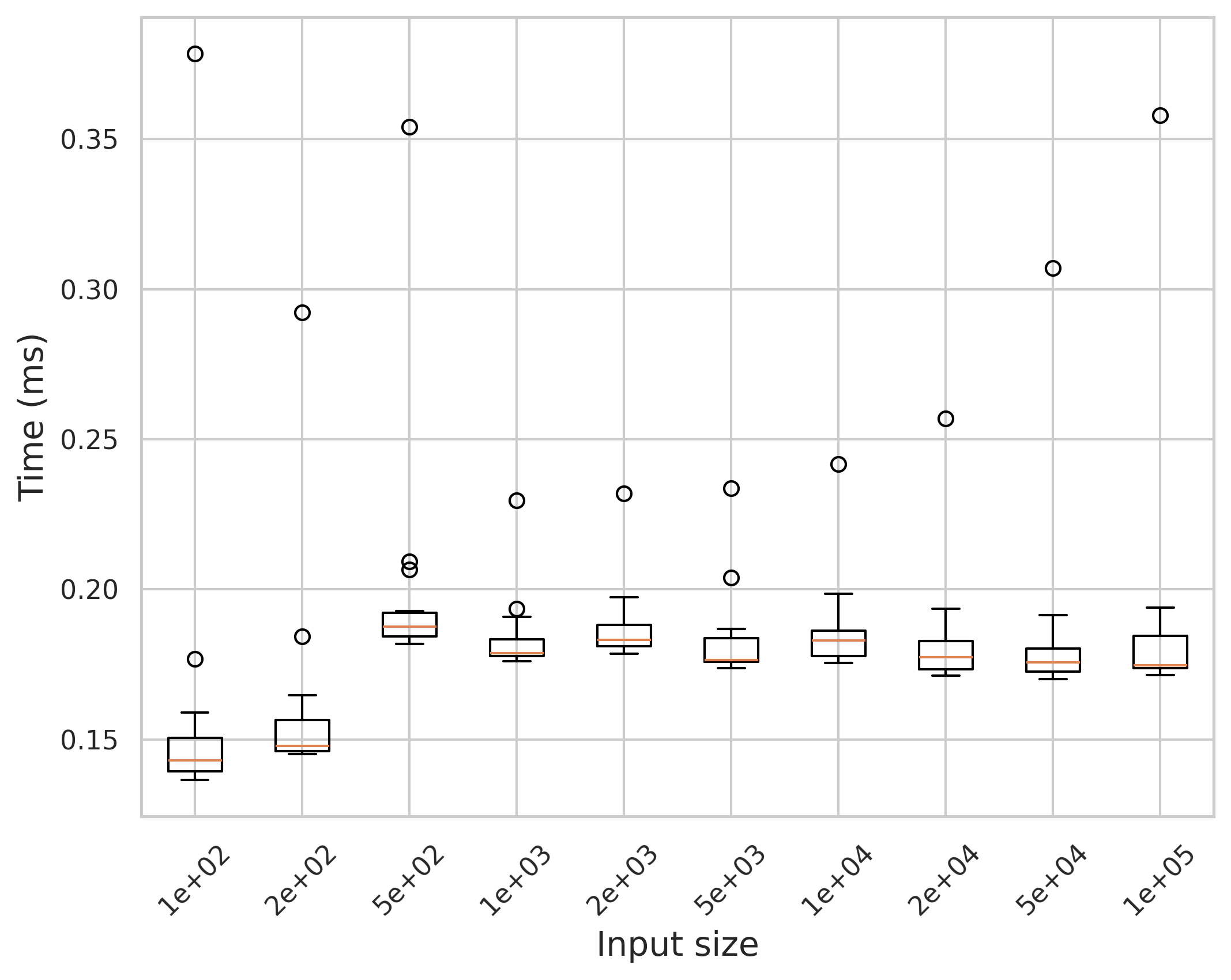}
 \caption{\label{fig:t_insize}Box plot showing waveform generation time as a function of input (batch) size. The orange line within each box represents the median time. The box itself shows the interquartile range (IQR), while the whiskers extend to the minimum and maximum times within 1.5 times the IQR\@. Outliers are shown as individual circles}
 
\end{figure}
To test the runtime performance of \dansur~ models, we generate $1\e{7}$ randomly sampled waveforms in batches of $1\e{5}$ using an NVIDIA H100 GPU.~Note that in this process we are now including the scaling from geometric to physical units. Figure~\ref{fig:t_cumul} shows the cumulative time taken by~\dansur~to generate the waveforms. We must note here that there is a warm-up period, typically around 500 milliseconds of overhead, attributable to CUDA initialization. However, once initialized, the total generation time for $10^7$ waveforms is achieved in under 40 milliseconds. 

Figure~\ref{fig:t_insize} presents a box plot of the generation time for sets of 20 batches of size $10^2$ to $10^5$. As may be expected, as long as GPU memory is not exceeded, the batch size does not significantly affect the mean generation time. This means that approximants in the vein of~\dansur~can be used to generate massive GW template banks in very short time frames. 

\subsection{Parameter estimation\label{sec:pe}}

\begin{figure*}
 \centering
 \includegraphics[width=1\linewidth]{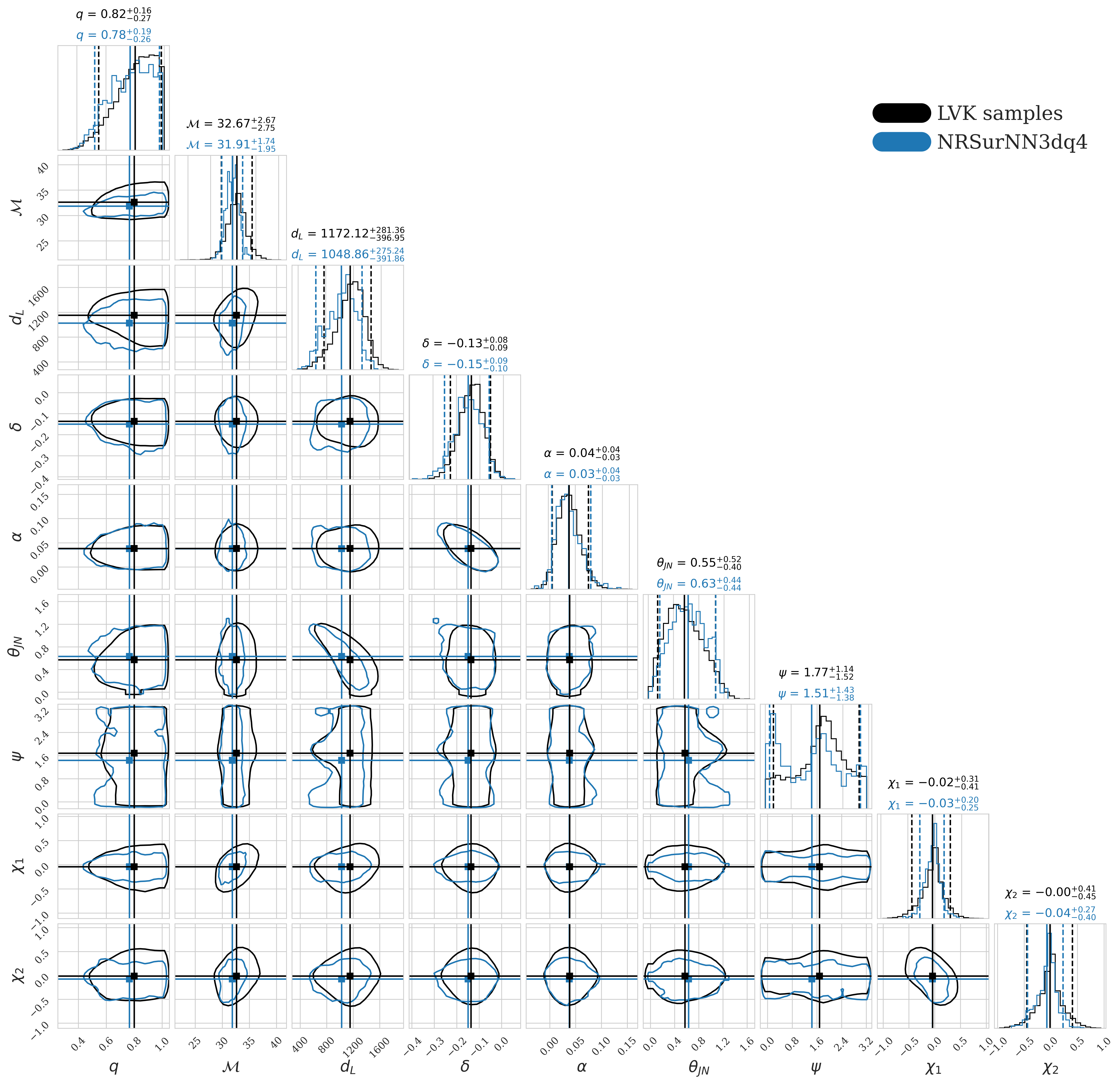}
 \caption{\label{fig:bilby_dists_full}Corner plot showing the relevant subset of posterior distributions of the~\textsc{bilby} parameter estimation. The dashed lines in the 1D histograms represent the 90\% credible intervals, while the isolines in the 2D plots represent the 90\% credible regions.}
 
\end{figure*}

As a way of exploring the ability of~\dansur~to be used in parameter estimation pipelines, we here discuss a basic test we conducted using the~\textsc{bilby} software suite for GWs~\cite{bilby_paper}. We must note that the current sampler and likelihood implementations in~\textsc{bilby} do not allow us at this point to take advantage of CUDA parallelization of the likelihood (though it is in principle possible to write a custom sampler that would manage this). Therefore, we will be running the model in the CPU for this section, focusing mostly on the accuracy of the model for parameter estimation.  To this end, we make use of \textsc{pytorch} (v2.4.1) model serialization through \textsc{TorchScript}, allowing for better performance and ease of use. For comparison's sake,  running the model on CPU,  implemented through the \textsc{gwsurrogate} api,  it takes us 0.3~s on average (over 10 runs) to generate 1000 waveforms in serial. On the other hand, in the same machine, it takes us 5.6~s seconds on average to generate 1000 waveforms with the ~\targsur~implementation in \textsc{gwsurrogate} (v1.1.6). We have restricted the generation with~\targsur~ to the (2,2) mode, and kept the same time grid as~\dansur. \change{These timing tests have been run on an AMD EPYC 9454 48-core CPU at the Artemisa cluster of the University of Valencia.}
For this experiment, we will focus on performing a parameter inference run on the GW200311\_115853 event from the third Gravitational-Wave Transient Catalog~\cite{GWTC-3}, using the~\textsc{nessai} sampler~\cite{nessai} and the~\dansur~ pretrained on the \texttt{NRHybSur3dq8} approximant as the waveform model. This event was chosen as an illustrative example due to its relatively high SNR, but performance in other events can be expected to be similar. With 1000 live points, this run took 3 minutes and 35 seconds to complete, with 346,596 likelihood evaluations accounting for 15.8 seconds of that time. In Fig.~\ref{fig:bilby_dists_full} we show the results of the PE run, with the posteriors being calculated for the mass ratio $q$ and the spins $\chi_1,~\chi_2$, as well as the chirp mass $\mathcal{M} = \frac{(m_1 m_2)^{3/5}}{(m_1 + m_2)^{1/5}}$, luminosity distance $d_L$, the sky position angular coordinates (right ascension $\alpha$ and declination $\delta$), and the polarization angle $\Psi$. All estimated values are contained in the 90\% credible regions of the LVK posterior distributions, showing that our model can be used to reliably reconstruct physical parameters in line with LVK standards. These results do have the caveat that our model does not include higher harmonics, which are more important the larger the mass ratio is, as well as allowing for better precision at higher SNR values. The lack of precession modeling must also be taken into account. This will bias our model under certain conditions.

\section{Conclusions}~\label{sec:conclusions}

In this paper we have introduced a 2-step training strategy to generate neural-network-based NR surrogate models (\dansur), shown to rapidly and accurately generate non-precessing BBH waveforms.

\change{
In a first step, we use BBH approximants to generate large datasets of gravitational waveforms for four different approximants, using the resulting datasets to pre-train surrogate networks. We show that, after the pretraining step, the resulting model can already accurately reproduce the outputs of the base approximant. Together with our usage of a reduced-order PCA basis, this step effectively restricts the vast function space of the neural network, so that when we start our second step, our initial point in the network's state space is one which produces gravitational waveforms from BBH systems. The second step itself consists of fine-tuning a pre-trained network on waveforms from NR simulations, allowing the training process to adjust the basis through backpropagation as well.  This results in median mismatches around $5\e{-5}$ on average, regardless of the approximant used for pretraining. When compared to the base approximant used for pre-training, we show that the fine-tuned \dansur~model can improve mismatches with NR waveforms by up to two orders of magnitude, reinforcing the effectiveness of the 2-step training procedure. }

Timing tests demonstrate that surrogates trained under this strategy can generate over a million waveforms in under a tenth of a second when implemented in a GPU. This makes them a powerful tool when the generation of large amounts of waveforms in parallel is required, such as building template banks, building datasets for training machine learning models, or for GPU-parallelized parameter estimation.

\change{Despite the studied case's limitations, namely in the absence of higher modes and precession, as well as the limited number of cycles used, the evaluation of the \dansur~model within the~\textsc{bilby} framework shows results that are consistent with those reported in the literature. The results for parameter estimation confirm the model's accuracy and open the door for further integration. A GPU adaptation of existing~\textsc{bilby} samplers and likelihood implementations in order to leverage the ability to generate arbitrarily large numbers of waveforms has the potential to speed up parameter estimation in a very significant way and must be the focus of future work. Additional future efforts will focus on augmenting \dansur's capabilities to model precessing systems and higher modes, refining its accuracy through the exploration of alternative dimensionality reduction methods and alternative network architectures, and extending the parameter space.}

\vspace{2cm}
\begin{acknowledgments}
The authors acknowledge the anonymous reviewer for their insightful and productive comments which have greatly helped in increasing the rigor of this work.
OGF is supported by the Portuguese Foundation for Science and Technology (FCT) through doctoral scholarship UI/BD/154358/2022.
TF is supported by FCT through doctoral scholarship (reference 2023.03753.BD).
OGF, TF, SN and AO acknowledge financial support by CF-UM-UP through Strategic Funding UIDB/04650/2020.
JAF and ATF are supported by the Spanish Agencia Estatal de Investigaci\'on (PID2021-125485NB-C21) funded by MCIN/AEI/10.13039/501100011033 and ERDF A way of making Europe. 
Further support is provided by the Generalitat Valenciana (CIPROM/2022/49) and  by  the  European Horizon  Europe  staff  exchange  (SE)  programme HORIZON-MSCA-2021-SE-01 (NewFunFiCO-101086251).
AO is partially supported by FCT, under the Contract CERN/FIS-PAR/0037/2021.
JDMG is partially supported by the agreement funded by the European Union, between the Valencian Ministry of
Innovation, Universities, Science and Digital Society, and the network of research centers in Artificial Intelligence
(Valencian Foundation valgrAI), as well as the Valencian Government grant with reference number CIAICO/2021/184; the Spanish Ministry of Economic Affairs and Digital Transformation through the QUANTUM ENIA project call – Quantum Spain project, and the European Union through the Recovery, Transformation and Resilience Plan – NextGenerationEU within the framework of the Digital Spain 2025 Agenda.
The authors gratefully acknowledge the computer resources at Artemisa and the technical support provided by the Instituto de Fisica Corpuscular, IFIC (CSIC-UV). Artemisa is co-funded by the European Union through the 2014-2020 ERDF Operative Programme of Comunitat Valenciana, project IDIFEDER/2018/048. This material is based upon work supported by NSF's LIGO Laboratory which is a major facility fully funded by the National Science Foundation, as well as the Science and Technology Facilities Council (STFC) of the United Kingdom, the Max-Planck-Society (MPS), and the State of Niedersachsen/Germany for support of the construction of Advanced LIGO and construction and operation of the GEO600 detector. Additional support for Advanced LIGO was provided by the Australian Research Council. Virgo is funded, through the European Gravitational Observatory (EGO), by the French Centre National de Recherche Scientifique (CNRS), the Italian Istituto Nazionale di Fisica Nucleare (INFN) and the Dutch Nikhef, with contributions by institutions from Belgium, Germany, Greece, Hungary, Ireland, Japan, Monaco, Poland, Portugal, Spain. KAGRA is supported by Ministry of Education, Culture, Sports, Science and Technology (MEXT), Japan Society for the Promotion of Science (JSPS) in Japan; National Research Foundation (NRF) and Ministry of Science and ICT (MSIT) in Korea; Academia Sinica (AS) and National Science and Technology Council (NSTC) in Taiwan 
\end{acknowledgments}

\bibliographystyle{unsrt}
\bibliography{bibliography}

\end{document}